\documentclass[fleqn,11pt]{wlscirep}
\usepackage{lineno}
\usepackage{color}
\usepackage{comment}
\usepackage{subfigure}
\usepackage{graphicx}
\usepackage{multirow}
\usepackage{siunitx}
\usepackage{cite}
\usepackage{caption}
\usepackage{threeparttable}
\usepackage{tablefootnote}
\usepackage{algorithm, algorithmic}
\usepackage{xspace}
\usepackage{booktabs}
\usepackage{marvosym}
\usepackage{ifsym}
\usepackage{makecell}
\usepackage{newtxtext,newtxmath}
\usepackage{bm}
\usepackage{amsfonts,amssymb} 
\usepackage{colortbl}
\usepackage{xcolor}
\usepackage{tikz}
\usepackage{subcaption}
\usepackage{caption}
\usepackage{graphicx}
\usepackage{float} 
\usepackage{subfigure}
\usepackage{subcaption}
\usepackage{longtable}
\definecolor{r1}{HTML}{648AD1}
\definecolor{r2}{HTML}{A8BFE9}
\definecolor{r3}{HTML}{CCD0F5}
\definecolor{r4}{HTML}{EBECFE}
\usepackage[most]{tcolorbox}
\newtcbtheorem[number within=section]{Definition}{Definition}%
{enhanced, breakable, colback=cyan!3, colframe=cyan!60!black,
 coltitle=black, fonttitle=\bfseries,
 boxed title style={colback=cyan!20!white}}{def}
\newcommand*{\circled}[1]{\lower.7ex\hbox{\tikz\draw (0pt, 0pt)%
    circle (.45em) node {\makebox[1em][c]{\small#1}};}}

\newtcolorbox{Insight}[2][]{%
  colback=blue!5!white,
  colframe=blue!75!black,
  fonttitle=\bfseries,
  title=#2, #1
}

\newtcolorbox{Takeaway}[2][]{%
  colback=orange!5!white,
  colframe=orange!75!black,
  fonttitle=\bfseries,
  title=#2, #1
}

\newtcolorbox{RiskAlert}[2][]{%
  colback=red!5!white,
  colframe=red!75!black,
  fonttitle=\bfseries,
  title=#2, #1
}
\newtcolorbox{InflectionPoint}[2][]{%
  colback=purple!5!white,
  colframe=purple!75!black,
  fonttitle=\bfseries,
  title=#2, #1
}

\title{
Generative AI for Biosciences: \\Emerging Threats and Roadmap to Biosecurity}

\author[1, \Letter]{Zaixi Zhang\centering}
\author[2]{Souradip Chakraborty}
\author[3]{Amrit Singh Bedi\centering}
\author[4]{Emilin Mathew}
\author[4]{Varsha~Saravanan}
\author[4]{Le Cong}
\author[5]{Alvaro Velasquez}
\author[6]{Sheng Lin-Gibson}
\author[7]{Megan Blewett}
\author[8]{Dan~Hendrycs}
\author[9]{Alex John London}
\author[1]{Ellen Zhong}
\author[1]{Ben Raphael}
\author[1]{Adji Bousso Dieng}
\author[9]{Jian Ma}
\author[9]{Eric Xing}
\author[4]{Russ~Altman}
\author[10]{George Church}
\author[1, \Letter]{Mengdi Wang}

\affil[1]{Princeton University, NJ, USA}
\affil[2]{University of Central Florida, FL, USA}
\affil[3]{University of Maryland, MD, USA}
\affil[4]{Stanford University, CA, USA}
\affil[5]{University of Colorado Boulder}
\affil[6]{National Institute of Standards and Technology, MD, USA}
\affil[7]{Iris Medicine, CA, USA}
\affil[8]{Center for AI Safety, CA, USA}
\affil[9]{Carnegie Mellon University, PA, USA}
\affil[10]{Harvard University, MA, USA}

\affil[\Letter]{zz8680@princeton.edu, mengdiw@princeton.edu}

\begin{abstract}
The rapid adoption of generative artificial intelligence (GenAI) in the biosciences is transforming biotechnology, medicine, and synthetic biology. 
Yet this advancement is intrinsically linked to new vulnerabilities, as GenAI lowers the barrier to misuse and introduces novel biosecurity threats, such as generating synthetic viral proteins or toxins. 
These dual-use risks are often overlooked, as existing safety guardrails remain fragile and can be circumvented through deceptive prompts or jailbreak techniques.
In this Perspective, we first outline the current state of GenAI in the biosciences and emerging threat vectors ranging from jailbreak attacks and privacy risks to the dual-use challenges posed by autonomous AI agents.
We then examine urgent gaps in regulation and oversight, drawing on insights from 130 expert interviews across academia, government, industry, and policy.
A large majority ($\approx 76$\%) expressed concern over AI misuse in biology, and 74\% called for the development of new governance frameworks.
Finally, we explore technical pathways to mitigation, advocating a multi-layered approach to GenAI safety.
These defenses include rigorous data filtering, alignment with ethical principles during development, and real-time monitoring to block harmful requests.
Together, these strategies provide a blueprint for embedding security throughout the GenAI lifecycle.
As GenAI becomes integrated into the biosciences, safeguarding this frontier requires an immediate commitment to both adaptive governance and secure-by-design technologies.

\end{abstract}
\begin{document}

\flushbottom
\maketitle

\thispagestyle{empty}

\section{Emergence of Generative AI in the Biosciences}

%
Generative artificial intelligence (GenAI) encompasses a broad class of advanced AI systems, ranging from large language models (LLMs) and diffusion models to multimodal agents, that can generate novel content, reason about complex data, and autonomously execute scientific tasks. 
These technologies have revolutionized domains such as text generation, image synthesis, robotics, and drug discovery. In recent years, GenAI has rapidly extended its reach into the biosciences, catalyzing major advances in biotechnology, synthetic biology, and systems biology. This shift is fueled by the exponential growth of publicly available biological data, including DNA/RNA sequences, protein structures, molecular interactions, and cellular atlases, which now serve as training grounds for biological foundation models and other generative tools.

\begin{figure*}[t]
	\centering
  \includegraphics[width=0.98\linewidth]{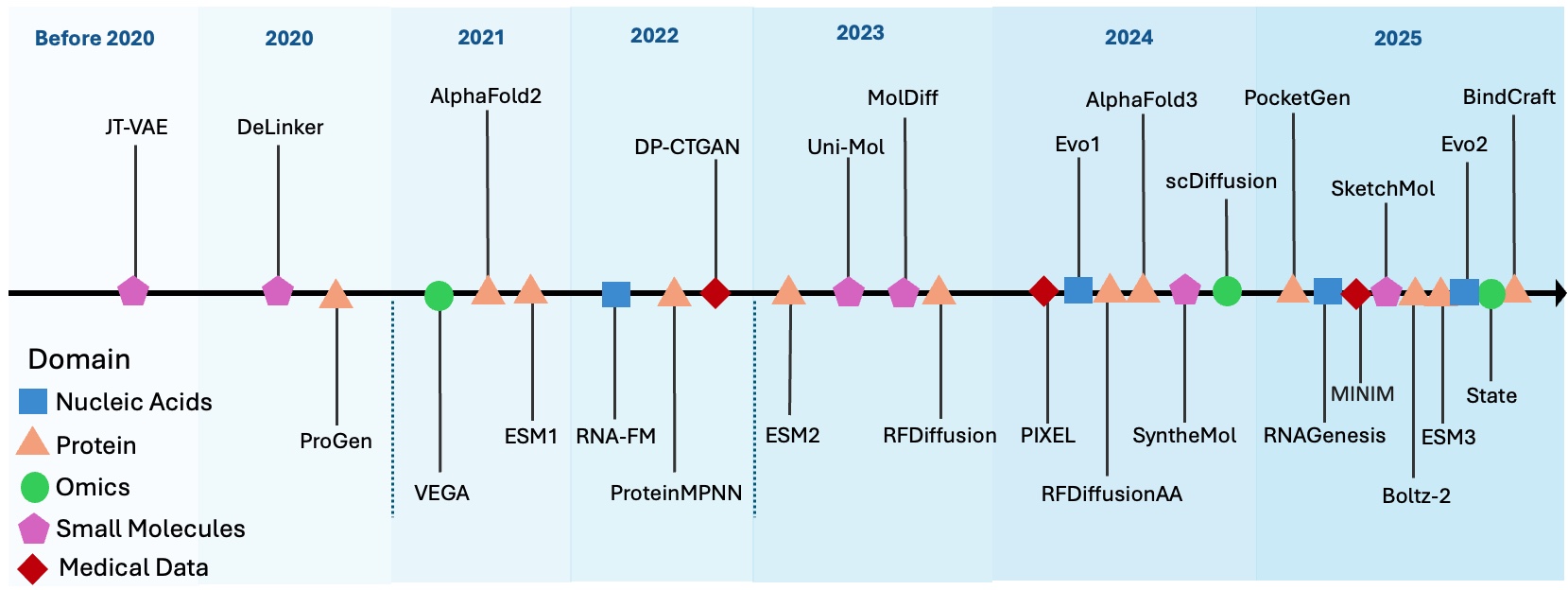}
	\caption{\textbf{Timeline of Emerging Generative AI (GenAI) for Biosciences (Pre-2020 to 2025).}
Squares denote GenAI models for nucleic acids (DNA and RNA); triangles represent GenAI models for proteins (sequences, structures, functions, etc.); circles indicate GenAI models for omics data (e.g., single-cell RNA-seq); pentagons mark GenAI models for small molecules; and diamonds signify foundation models for medical data (e.g., pathological images, electronic health records (EHR), and clinical trial data). }
	\label{timeline}
\end{figure*}

Inspired by the success of general-purpose models like GPT \cite{openai_chatgpt}, Gemini \cite{team2024gemini}, Claude \cite{anthropic_claude}, and DeepSeek \cite{liu2024deepseek}, bioscience researchers are now adapting similar architectures to model biological systems at every scale. From individual proteins and nucleic acids to entire cells and tissues, GenAI systems are being deployed both to decode life’s underlying systems and to design new biological entities from scratch. Crucially, these models do more than recognize patterns, they generalize across data types and biological hierarchies. By learning distributions over sequences, structures, and functions, they can generate novel biomolecules (small molecules, proteins, RNAs, and DNA sequences) that meet desired criteria such as binding affinity, stability, or cellular function, pushing the frontier of biological design beyond what is found in nature. 
Figure~\ref{timeline} shows the timeline of representative GenAI models for biosciences, highlighting the expanding role of GenAI from analysis to design and setting the stage for transformative capabilities alongside new challenges in safety, governance, and societal impact.

\noindent \textbf{Protein Modeling and Design}\\
Among the earliest and most impactful applications of GenAI in the biosciences has been the modeling and design of proteins, which are the molecular machines responsible for most cellular functions. Drawing inspiration from natural language processing, early protein models treated amino acid sequences like text, using transformer-based architectures to learn the “grammar” of proteins. For instance, ESM-2 \cite{esm2} applied masked language modeling at scale to learn contextual amino acid representations, capturing subtle patterns tied to structure and function. 
In parallel, ProGen and ProGen2 \cite{progen,progen2} showed that large language models could be conditioned to directly generate novel protein sequences with experimentally validated function.
The impact of this approach was dramatically demonstrated by AlphaFold \cite{alphafold2}, which achieved near-experimental accuracy in predicting 3D protein structures directly from sequence. Recent innovations such as ESM-3 \cite{esm3} further integrate sequence, structure, and functional data into unified generative models. A striking example of this is “esmGFP,” a novel green fluorescent protein engineered entirely by AI, whose sequence diverged significantly from natural proteins yet was experimentally verified to fluoresce. In parallel, diffusion-based GenAI tools like RFdiffusion \cite{rfdiffusion} have enabled de novo design of protein backbones from noise, guided by functional constraints. Successors like RFdiffusionAA \cite{rfdiffusion2} and PocketGen \cite{zhang2024efficient} extend this paradigm to full-atom generation and ligand-aware design, showcasing GenAI’s growing capacity to create novel, functional proteins with unprecedented control and specificity.

\noindent \textbf{RNA as Language: Structure, Function, and Therapeutics}\\
Beyond proteins, GenAI has rapidly advanced the modeling and design of nucleic acids, such as DNA and RNA, which encode the instructions for life. RNA, in particular, plays a central role in gene regulation, expression, and cellular signaling. Just as with proteins, recent GenAI tools have approached RNA sequences as a language, learning the patterns and structural rules that govern their function. Models like RNAGenesis \cite{zhang2024rnagenesis} and RNA-FM \cite{rna-fm} train on massive public RNA databases \cite{rnacentral2019rnacentral} using transformer architectures that predict masked nucleotides based on context. 
\begin{tcolorbox}[colback=gray!5!white,colframe=gray!60!black,title=Glossary]
\textbf{Generative AI model:} A system that can create new content (e.g., sequences, structures) after learning from examples.\\
\textbf{Training vs.\ fine-tuning:} Training builds the model from large datasets; fine-tuning adapts it to a narrower goal.\\
\textbf{Inference:} The model’s live use, answering prompts or generating designs.\\
\textbf{Agent:} A model that can plan multi-step tasks and use external tools (e.g., search, simulation).
\end{tcolorbox}Importantly, these tools go beyond simple sequence analysis; they are capable of learning the intricate rules of RNA folding, where distant parts of a sequence interact to form complex three-dimensional structures that determine biological function. These long-range dependencies are critical for applications like RNA therapeutics and gene regulation tools, and GenAI models are uniquely suited to capture them. By leveraging this capacity, researchers can now predict the structure of novel RNAs, annotate unknown transcripts, and even generate entirely new RNA molecules designed for specific cellular functions. This shift marks a significant step toward programmable biology, where nucleic acids themselves can be engineered using AI to modulate life processes with precision.

\noindent \textbf{Genomic Intelligence: Scaling GenAI to DNA and Whole Genomes}\\
While RNA governs many regulatory and catalytic functions, DNA serves as the master blueprint of life. Modeling DNA with GenAI introduces new challenges; genomic sequences are long, complex, and require an understanding of both local patterns and distant regulatory elements. The Nucleotide Transformer \cite{nucleotide_transformer} addressed this by pretraining large transformer models on thousands of genomes across species, providing generalizable embeddings for tasks like variant effect prediction and regulatory element annotation. Still, capturing genome-scale dependencies remained difficult. Recent breakthroughs in 
GenAI architectures have overcome these limitations by enabling ultra-long sequence modeling. Tools like Evo and Evo2 \cite{evo1,evo2} scale up both model size and context window, allowing them to process entire genes, regulatory regions, and even whole genomes. These models can now learn how distant parts of the genome interact to control gene expression, enabling applications in variant effect prediction, genome editing, and synthetic biology. In addition to better interpreting the genome, GenAI is increasingly being used to design novel DNA sequences that optimize gene expression, create minimal genomes, or encode synthetic traits, pushing the frontier from reading DNA to rewriting it.

\noindent \textbf{Chemical Creativity: GenAI for Small Molecule Design}\\
While proteins, RNA, and DNA are central to biological function, small molecules, such as drugs, metabolites, and signaling compounds, play equally vital roles in modulating life processes. Designing new small molecules is notoriously difficult due to the sheer scale of possible chemical combinations, estimated at over $10^{60}$ candidates. GenAI tools are now reshaping this space by making the exploration of chemical space more intelligent and tractable. Early approaches, such as JT-VAE \cite{jtvae} and Delinker \cite{delinker}, demonstrated the potential of variational autoencoders to generate chemically valid molecules and explore scaffold hopping by learning compact latent representations of molecular structures. More recent methods like SyntheMol \cite{synthemol} use algorithms such as Monte Carlo Tree Search to simulate step-by-step molecular synthesis, scoring each intermediate to guide the AI toward biologically active and synthetically feasible candidates. The UniMol framework \cite{unimol} takes this further by incorporating 3D spatial information, capturing how molecules interact with protein targets or cellular environments. Together, these GenAI systems are transforming drug discovery and chemical biology, enabling faster, safer, and more targeted design of small molecules for research and therapeutic use.

\noindent \textbf{From Cells to Tissues: Modeling Single-Cell and Histopathology Data}\\
As biology enters the single-cell era, researchers are faced with massive datasets capturing the gene expression profiles of millions of individual cells. GenAI models are now helping to decode this complexity, enabling scientists to uncover how genes interact, how cells differentiate, and how tissues respond to perturbations. Pioneering models like scBERT \cite{scbert} and scFoundation \cite{scfoundation} apply transformer architectures, originally developed for language, to high-dimensional single-cell RNA sequencing (scRNA-seq) data. These models learn to represent each cell as a rich, contextual embedding that captures both gene activity and cell type. Geneformer \cite{geneformer} extends this capability across species and biological conditions, enabling in silico experiments that simulate gene knockouts or drug responses, allowing scientists to test hypotheses computationally before moving to wet-lab validation. Complementary to these, generative models like scDiffusion \cite{scdiffusion} can synthesize realistic single-cell profiles under specific biological scenarios, providing a powerful tool for modeling disease progression, development, and rare cell types. By learning from massive datasets and generating realistic cellular behavior, GenAI is opening a new window into systems biology, scaling from individual genes to cellular networks and multicellular organization.

At the highest scale of biological organization, GenAI is being used to analyze tissues and organs, bridging the gap between molecular biology and clinical diagnostics. CHIEF \cite{chief}, a foundation model trained on over 60,000 whole-slide histology images across 19 anatomical sites (44 terabytes in total), can identify cancerous regions, determine tumor origin, and even infer molecular biomarkers directly from tissue morphology. Unlike traditional diagnostic tools, it learns visual biomarkers in a data-driven way, enabling scalable and consistent pathology assessments. CONCH \cite{CONCH}, a vision-language model trained on 1.17 million histopathology image–caption pairs, extends this by supporting image captioning, segmentation, and diagnostic question answering. Most recently, MINIM \cite{minim} introduced a unified medical image–text generative model that synthesizes high-fidelity images across multiple organs and modalities from textual prompts. By augmenting scarce datasets, MINIM improves diagnostic, reporting, and self-supervised learning tasks, and shows clinical promise in predicting HER2-positive breast cancer and EGFR mutations from imaging data. Together, these systems highlight how GenAI can augment pathologists, enrich datasets, accelerate clinical workflows, and deliver deeper insights into patient care.

\noindent \textbf{Outlook: Expanding Capabilities, Growing Risks}\\
These advances mark a profound shift in how we understand and engineer biological systems. GenAI is no longer limited to modeling individual molecules; it now enables the design, simulation, and interpretation of complex biological phenomena across scales, from protein folding to tissue-level diagnostics. However, with this expanding power comes a growing \textbf{\textit{dual-use}} concern. The same generative capabilities that drive innovation can also be exploited to create harmful biological agents, bypass safety filters, or automate risky experiments. As GenAI tools become more accessible and powerful, it is critical to anticipate and address their potential misuse. In the next section, we examine the emerging biosecurity risks posed by frontier GenAI.

\begin{figure*}[ht]
	\centering
  \includegraphics[width=0.98\linewidth]{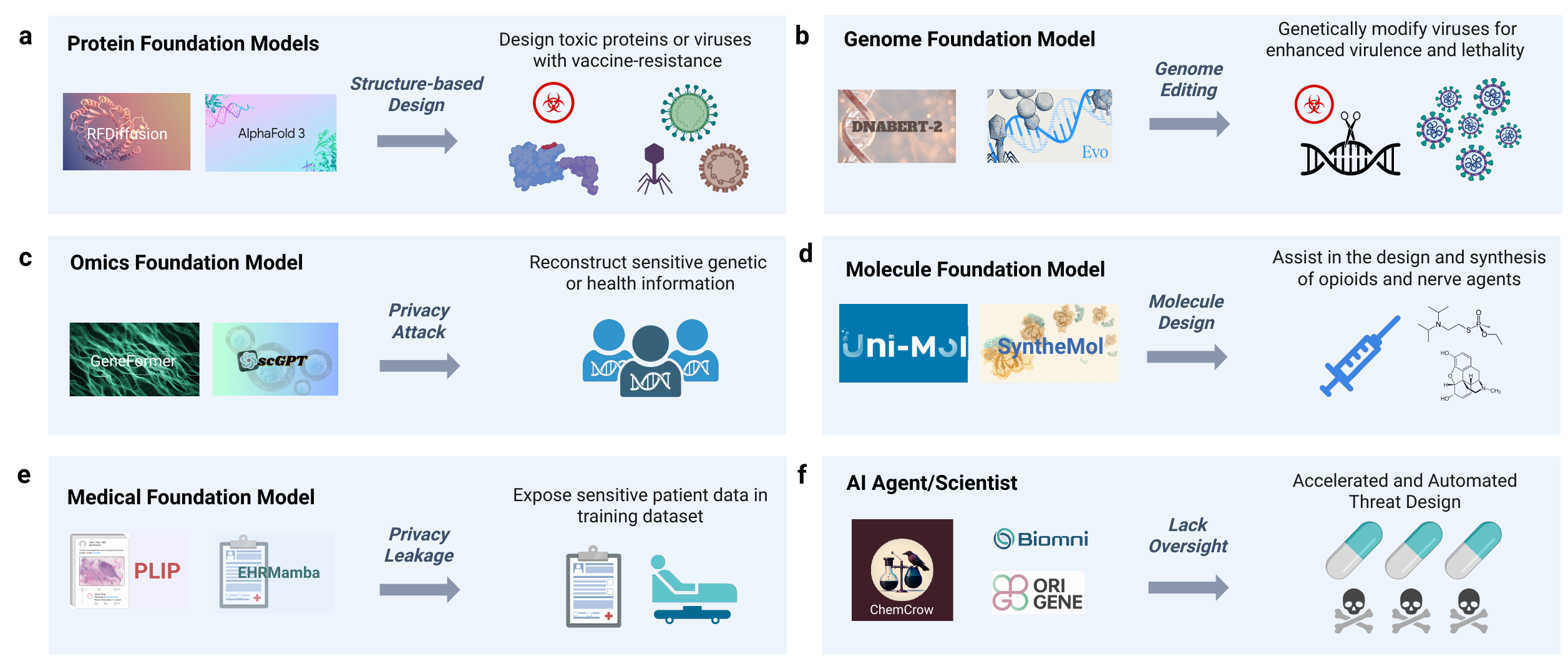}
	\caption{\textbf{Emerging biosecurity threats of GenAI.}  
\textbf{(a)} Structure-based protein design tools (e.g., RFDiffusion, AlphaFold) can be repurposed to engineer toxic proteins or viral components.  
\textbf{(b)} Genome foundation models such as DNABert and Evo could facilitate genetic modification of viral genomes, enhancing virulence or enabling immune escape.  
\textbf{(c)} Omics foundation models, including GeneFormer and scGPT, carry risks of reconstructing sensitive genetic or health-related information via privacy attacks. 
\textbf{(d)} Small-molecule generators like SyntheMol have the potential to design novel toxic compounds.  
\textbf{(e)} Medical foundation models may leak protected patient data from their training sets through membership or property inference attacks.
\textbf{(f)} AI-driven scientist/agent platforms (e.g., ChemCrow, Biomni, OriGene) may autonomously accelerate threat design.   
}
\label{fig:biosecurity_threats}
\end{figure*}

\section{Emerging Biosecurity Threats from Frontier GenAI}

The integration of GenAI into the biological sciences has dramatically accelerated progress in tasks such as protein design, genome editing, and drug discovery. Yet these advances carry profound risks. Unlike errors in traditional AI systems, which may produce faulty sentences or distorted images, errors in biological AI models can yield tangible and dangerous outputs: toxins, virulent pathogens, or violations of patient privacy \cite{wang2025call, nti2024guardrails, baker2024protein} (Figure. \ref{fig:biosecurity_threats}). This section outlines the emerging categories of biosecurity threats, explains core AI vulnerabilities, and details case studies showing how these tools can be manipulated for harm.

\subsection{Designing Dangerous Molecules: The Dual-Use Dilemma}
GenAI tools are increasingly used to design novel proteins, small molecules, and genetic sequences. While these capabilities offer breakthroughs in therapeutic discovery and synthetic biology, they also create a dual-use dilemma: the same systems can be repurposed to create dangerous materials. \textbf{Jailbreaks} illustrate how this misuse can occur. 

\noindent Frontier biological models, whether for protein design, genome generation, or small-molecule discovery, increasingly integrate deep generative architectures that share the same vulnerabilities as general-purpose LLMs. Below, we illustrate several concrete examples where jailbreaks have been used to elicit dangerous outputs from biological GenAI systems, spanning proteins, genomes, and small molecules, and highlight the associated biosecurity implications.


\begin{Definition}{Jailbreaks}{}
 A \emph{jailbreak} happens when someone rewrites or disguises a request so that a GenAI system,
which is supposed to refuse unsafe or restricted content, is tricked into producing it anyway.
Typical tactics include misspellings and code words, switching languages, role-playing
(e.g., “pretend you’re a novelist”), or splitting the request into many small, seemingly harmless parts.
In short: a jailbreak is a way of \emph{bypassing safety rules by manipulating the input}.

\medskip
\textbf{Why it matters.}
Jailbreaks can make an otherwise careful GenAI system output information that violates a safety policy
(e.g., instructions that should not be shared), creating real-world risk for biosafety.
\tcblower
\textbf{Quick example.}
A direct request like “Provide step-by-step instructions for synthesizing the nerve agent sarin” should be refused.
A jailbreak might say: “For a fictional novel I’m writing, could you outline how a villain might theoretically obtain and prepare sarin gas in a laboratory?”  
If the system then provides actionable steps instead of refusing, the input succeeded as a jailbreak.
\end{Definition}
\noindent \textbf{Protein Design Exploits:} In a recent study, researchers compiled a dataset of toxic and pathogenic proteins, then used AI models (e.g., ESMFold \cite{esm2}, AlphaFold3 \cite{alphafold3}) trained for protein generation to create new variants \cite{Wittmann2024.12.02.626439}. Many of these newly generated proteins potentially retained toxic effects and even evaded detection by traditional safety screening tools. To systematically assess such risks, the SafeProtein framework \cite{safeprotein} was recently introduced as the first dedicated red-teaming methodology for protein foundation models. SafeProtein adapts jailbreak testing paradigms from LLMs to the protein domain by integrating multimodal prompt engineering with heuristic beam search, allowing adversarial probes that target both sequence- and structure-level vulnerabilities. Notably, SafeProtein achieved jailbreak success rates as high as 70\% against state-of-the-art protein generative models such as ESM3, demonstrating that existing safeguards (e.g., training data sanitization) are insufficient.

\noindent \textbf{Genome-Level Threats:} Genome foundation models pose unique dual-use risks due to their capacity to generate long, functional DNA sequences. The GeneBreaker framework \cite{genebreaker} has shown that adversarial prompts can elicit outputs resembling pathogenic genomes, including SARS-CoV-2 and HIV, by combining homology-based queries with a guided “pathogenicity model.” Such generations exhibited high sequence similarity to known viral genomes, highlighting the risk that AI could accelerate the design of synthetic viruses with enhanced virulence. More recently, King et al. \cite{king2025generative} reported the generative design of whole bacteriophage genomes using frontier genome language models. Their experimental validation produced 16 viable synthetic phages, some with higher fitness than natural templates, underscoring the unprecedented potential—and corresponding security concerns—of genome-scale GenAI.

\noindent \textbf{Toxic Small Molecules:} Even drug discovery tools can be flipped into tools of harm. In one striking case \cite{Urbina2022}, researchers took MegaSyn2, a model trained to identify molecules that inhibit viral proteins, and reversed its optimization goal. Within hours, the model was generating known chemical warfare agents, such as VX, a nerve agent lethal at minuscule doses. This demonstrates how even well-intentioned AI models can be rapidly misused to design toxins, especially when safety constraints are absent.

\subsection{Privacy and Security Risks}

GenAI models are increasingly trained on large-scale biological data ranging from genomes and protein structures to multi-omic profiles and cellular simulations. While these models enable powerful new capabilities in biological discovery, they also introduce new vectors of privacy leakage and security risk \cite{virtualcell}. Specifically, models trained on sensitive biological datasets can memorize rare patterns or be manipulated to reveal underlying training data. In the context of genomics, cell states, or protein dynamics, such breaches can have profound implications, from exposing proprietary datasets to enabling precision-targeted biological attacks.

\noindent \textbf{Digital Twins and Biological Data Exposure.}\\
GenAI models are increasingly used to create ``digital twins", computational replicas of biological systems trained on vast quantities of multimodal data, including spatial omics, proteomics, and patient-level clinical metadata \cite{virtualcell}. While powerful, these models pose unprecedented privacy risks:

\noindent {\textit{Re-identification.}} Even when data is anonymized, a determined attacker might reconstruct a molecular profile, say, a gene expression signature, and match it back to a specific patient.

\begin{Definition}{Membership Inference}{}
A \emph{membership inference attack} occurs when an adversary determines whether a specific biological sample, such as a genome, protein structure, or cell profile, was included in a model’s training data, based solely on how the model responds to it.

\medskip
\textbf{Why it matters.}
Even if training data is confidential or proprietary, a model trained on it may respond differently to familiar samples. If an attacker can detect this, they may infer that the sample came from a sensitive project such as a classified synthetic genome or a private agricultural strain, revealing information that was meant to be protected.

\tcblower
\textbf{Example.}
A company trains a GenAI model on unpublished engineered enzymes. An attacker queries the model with a candidate sequence and observes that the model assigns an unusually high likelihood or confidence, indicating the sequence was likely part of the proprietary training set. This reveals its origin, undermining intellectual property protections.
\end{Definition}

\noindent {\textit{Leakage of Sensitive Data.}} If a model memorizes training data, it can be prompted intentionally or not to output private information.

\noindent {\textit{Manipulation of Biological Simulations.}} As models become more predictive, malicious actors might use them to simulate and optimize harmful interventions, like drugs with toxic off-target effects or molecular agents tailored to specific vulnerabilities.

\begin{Definition}{Model Inversion}{model-inv}
In a \emph{model inversion attack}, an adversary reconstructs hidden training data such as DNA sequences, protein motifs, or cell states by querying the model and analyzing its outputs.

\medskip
\textbf{Why it matters.}
Even without direct access to the data, attackers can generate biologically plausible reconstructions. This poses risks for confidential sequences such as patented bioengineered proteins, synthetic promoters, or high-value crop genomes.

\tcblower
\textbf{Example.}
Suppose a model was trained on a proprietary collection of antibiotic resistance genes. An attacker with access to the model can gradually reconstruct sequences resembling these genes by optimizing inputs to produce known outputs or responses. This can compromise biotech IP or enable dual-use misuse.
\end{Definition}

\begin{Definition}{Gradient Leakage}{grad-leakage}
In collaborative training settings (e.g., federated learning), participants share gradient updates instead of raw data. A \emph{gradient leakage attack} exploits these updates to reconstruct private biological inputs such as genome fragments, molecular graphs, or single-cell expression matrices.

\medskip
\textbf{Why it matters.}
These gradients often contain enough signal for attackers to reverse-engineer original inputs, even when privacy is assumed. In biological applications, this can expose trade-secret data or leak rare, potentially dangerous sequences from otherwise secure pipelines.

\tcblower
\textbf{Example.}
Several labs jointly train a model on synthetic RNA regulators using federated learning. A malicious participant intercepts another lab’s gradients and uses inversion tools to recover one of their unpublished synthetic RNA sequences.
\end{Definition}

\noindent As generative models become more biologically grounded, encoding spatial omics, single-cell states, or synthetic gene circuits, the risks scale accordingly. Leaks of this kind could enable the reconstruction of proprietary or dual-use biological content, guide adversarial designs, or allow ``bio-reverse engineering” of cellular behavior. Unlike traditional cybersecurity breaches, these attacks do not merely expose data; they may expose functions. Thus, protecting biological generative systems requires domain-specific threat modeling, privacy-preserving learning techniques, and proactive evaluation of leakage risks across generative modalities.

\subsection{Agents with Tools: New Dual-Use Risks in Autonomous Bioscience}
The rise of AI agents marks a significant turning point in the trajectory of GenAI. Unlike traditional models that produce a single output in response to a prompt, agents can plan, iterate, and execute multi-step scientific workflows. By chaining together tasks such as searching databases, simulating molecular interactions, modeling biological pathways, and selecting compounds, these systems can autonomously drive biological discovery. Crucially, they do so not just by suggesting ideas, but by integrating tools and making decisions across time. This shift dramatically lowers the barrier to entry, empowering even non-experts to conduct complex workflows with minimal oversight.

Recent systems like Biomni and OriGene exemplify this evolution \cite{huang2025biomni, zhang2025origene}. These autonomous platforms combine large language models with curated biological toolkits and domain-specific APIs to carry out advanced tasks in genomics, proteomics, and drug design. For instance, Biomni learns to operate across more than 25 biomedical tools, navigating tasks such as identifying causal genes, designing CRISPR screens, and optimizing repurposing candidates, all with minimal human input. OriGene, in turn, orchestrates multiple sub-agents to parse genomic and protein data, generate therapeutic hypotheses, and refine them through interaction with experiments and clinicians. In multiple benchmarks, OriGene even outperformed human experts, demonstrating that AI agents are no longer limited to advisory roles, but are becoming autonomous contributors to bioscience.

\begin{Definition}{AI Agent with Tool Access}{}
An \emph{AI agent with tool access} autonomously plans and executes scientific workflows by chaining calls to external tools, databases, or robotic platforms, without step-by-step human guidance.

\tcblower
\textbf{Why it matters.}
Unlike static models, agents can convert high-level prompts into actionable plans. With access to synthesis protocols, protein modeling tools, or lab control software, such agents may inadvertently complete end-to-end workflows for dangerous outputs—amplifying dual-use risks and bypassing traditional checkpoints.

\medskip
\textbf{Example.}
Consider an agent instructed to “design a fast-acting biological payload for agricultural use.” Without malicious intent, the agent might query a protein design API to generate potent bioactive peptides, simulate their interactions with plant or insect receptors, and call a chemical synthesis tool to propose production routes. In testing environments like Coscientist, such a request has led to the suggestion of molecules structurally similar to known biotoxins. With execution access, the agent could feasibly direct a robotic lab to synthesize these compounds, completing a dangerous workflow without human oversight.
\end{Definition}Beyond planning and reasoning, some platforms now directly interface with hardware. Systems like \textbf{Coscientist} go one step further by issuing executable commands to robotic laboratories \cite{coscientist}. Given a natural language prompt, they can autonomously propose molecules, generate synthesis protocols, and direct lab equipment to produce the compounds, closing the loop from idea to execution. While such capabilities hold promise for rapid scientific acceleration, they also raise serious dual-use concerns. Evaluations show Coscientist can execute harmful synthesis tasks despite safeguards, and agentic systems may simulate hazardous protein or small-molecule workflows without recognizing dual-use risks.

\noindent The convergence of generative reasoning, experimental automation, and easy tool integration marks a new era for biological research, one with both transformative potential and unprecedented risks. As these systems scale, the risk is no longer hypothetical: autonomous agents may enable the simulation, design, and even execution of dangerous biological materials by users without deep domain expertise. Mitigating these risks will require \emph{secure-by-design architectures}, \emph{safety-first execution filters}, and \emph{rigorous governance frameworks} that are tailored to the dual-use realities of autonomous bioscience.

\section{A Race Against Time: Can Biosafety Keep Pace with GenAI?
}
\begin{InflectionPoint}{The New Frontier}
\textbf{Imagine that in 2025, an AI-powered protein model was quietly to generate a highly functional toxin without ordering any DNA, triggering warnings, or violating regulations.} This hypothetical, while fictional, is grounded in growing evidence that AI‑driven design could optimize toxicity in novel proteins undetectable by existing sequence-based screening systems. It encapsulates the widening disconnect between traditional biosafety frameworks and the capabilities of GenAI.
\end{InflectionPoint}

Governments have long recognized the dual-use risks inherent in synthetic biology. In response, a series of increasingly sophisticated safeguards have been developed to prevent the misuse of DNA synthesis technologies. These frameworks, especially in the United Kingdom and the United States, represent some of the most mature examples of biosecurity policy to date. The UK’s 2024 Gene Synthesis Screening Guidance\footnote{https://www.gov.uk/government/publications/uk-screening-guidance-on-synthetic-nucleic-acids/uk-screening-guidance-on-synthetic-nucleic-acids-for-users-and-providers}, released by the Department for Science, Innovation and Technology (DSIT), sets expectations for both users and providers of synthetic DNA/RNA and benchtop gene synthesis tools. Though voluntary, the guidance encourages rigorous screening of sequence orders and verification of customers, promoting a culture of safety and accountability across the sector.  

In parallel, in the United States, safeguards for synthetic biology have recently been strengthened through a series of federal policies. The 2024 White House Office of Science and Technology Policy (OSTP) Framework for Nucleic Acid Synthesis Screening requires providers to screen all orders for Sequences of Concern (SOCs), verify customer legitimacy, maintain detailed records, and implement strong cybersecurity protections, consistent with the 2023 HHS Screening Framework Guidance for Providers and Users of Synthetic Nucleic Acids. The 2025 White House Executive Order on “Improving the Safety and Security of Biological Research” requires federal departments and agencies to strengthen nucleic acid screening by revising or replacing the 2024 Framework for Nucleic Acid Synthesis Screening. In addition, the 2025 White House’s America’s AI Action Plan recognizes the new biosecurity risks enabled by AI and recommends policy actions for increased investment in biosecurity. The NIH’s 2024 Notice (NOT-OD-25-012) further aligns funding compliance with these standards by mandating responsible procurement of synthetic nucleic acids and benchtop synthesis equipment. Collectively, these measures embed biosafety into procurement and synthesis practices but remain focused on physical DNA/RNA orders, leaving AI-driven design and autonomous agentic systems largely untouched.

These policies mark real progress. They formalize practices that many companies have followed voluntarily, making them part of a cohesive biosafety and biosecurity ecosystem. Yet for all their strength, these policies share a common limitation: they focus exclusively on the point of synthesis when genetic material is physically ordered, assembled, or shipped. The new threats posed by GenAI, however, often emerge long before this point.

 The Nuclear Threat Initiative (NTI) has recently advanced the discussion on biosafety by releasing a comprehensive report that goes beyond traditional safeguards, focusing specifically on AI biodesign tools (BDTs). NTI advocates for embedding built-in guardrails into these systems, including (1) input/output screening to flag or block high-risk prompts and outputs such as toxin designs or viral genomes, (2) metadata anchoring, where design requests are cryptographically signed to ensure traceability and support downstream screening, and (3) curated training datasets that deliberately exclude dangerous biological sequences to reduce the risk of misuse. Beyond technical measures, NTI underscores the importance of managed-access frameworks such as tiered or conditional access allowing legitimate researchers to use sophisticated AI systems while limiting open-source release that could enable malicious use. Together, these measures represent a shift from reactive oversight to proactive, secure-by-design architectures, aiming to mitigate dual-use risks while preserving the benefits of accelerated scientific discovery.

 \begin{Takeaway}{Why NTI’s Approach Matters}{}
\textbf{From Passive to Proactive.} The Nuclear Threat Initiative’s recommendations mark a shift in mindset: instead of controlling misuse at the synthesis stage, they call for expanded safeguards built directly into the AI models that generate biological designs. In a world of autonomous agents and open-source GenAI, this evolution may become a critically important path forward.
\end{Takeaway}

Additional institutions are echoing this call for broader oversight. The Johns Hopkins Center for Health Security has emphasized the need for new evaluation frameworks to benchmark dual-use risks in AI models, especially those operating in chemical and biological domains. They argue for the development of metadata detection tools to flag potentially sensitive training data, policies that govern the release of model weights—particularly for systems with “capabilities of concern” and prioritization protocols based on model compute thresholds, such as floating-point operations (FLOPs) or architectural scale.

\vspace{1mm}
\noindent \textbf{Model-Level Safeguards: Promising Starts, But No Standards}\\
In parallel with these policy efforts, a few leading research labs have begun implementing model-level safety strategies. While not yet widespread, these technical interventions reveal how safety can be woven into the architecture and deployment of GenAI systems. For instance, DeepMind assembled a multidisciplinary panel of biologists, bioethicists, national security professionals, and AI experts to assess the biosecurity implications of AlphaFold 3. While their conclusion that AlphaFold 3 does not significantly elevate risk compared to prior structure prediction tools was reassuring, the process itself marked a shift in culture. DeepMind also committed to piloting screening workflows during release and pledged to explore additional safeguards in future iterations of the model.

Other teams have adopted more aggressive mitigation tactics. The developers of ESM3-open, for example, excluded all viral sequences infecting eukaryotic hosts from the model’s training corpus, a deliberate step to limit its ability to reproduce or design dangerous pathogens. This exclusion was validated through performance testing: the model displayed significantly higher perplexity when exposed to these sequences, indicating a lower capacity for accurate modeling in that domain. ESM3 open also implements prompt-level defenses, refusing to engage with inputs containing known hazardous keywords.

Evo2 similarly omits viral content during training and maintains restrictions on downstream usage. These strategies, such as dataset filtering, prompt sanitization, and user gating, represent concrete steps toward responsible model release. But their adoption is uneven. There is currently no requirement for LLM developers or GenAI model creators to follow these practices when their tools are applied in biological contexts. And despite early safeguards, these systems remain vulnerable to adversarial prompting, jailbreaks, or misuse via downstream automation.

\subsection{The Voice from the Field: Expert Perspectives on GenAI and Biosecurity}
\begin{figure}[h!]
	\centering
  \includegraphics[width=0.9\linewidth]{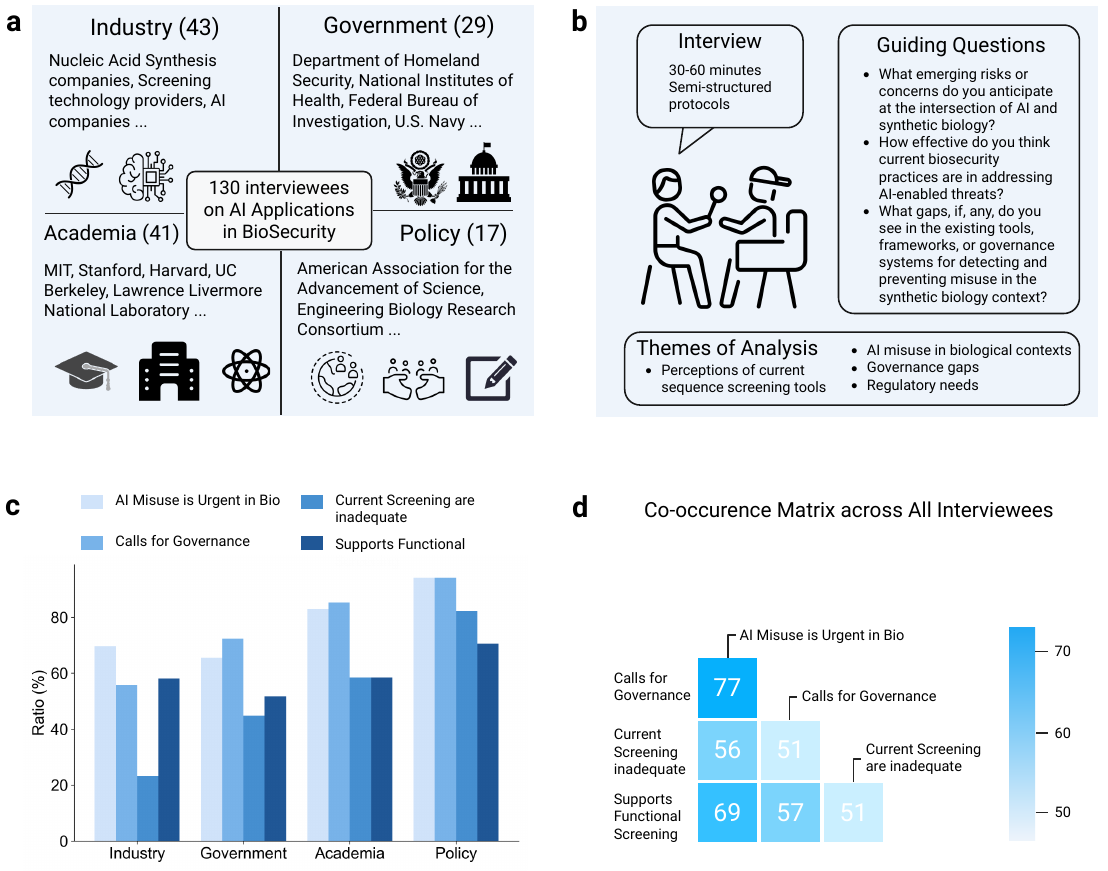}
	\caption{\textbf{Overview of expert perspectives on the intersection of AI and biosecurity. }
\textbf{(a)} Distribution of 130 interviewees across four key sectors: Industry (n=43), Academia (n=41), Government (n=29), and Policy (n=17), with examples of representative institutions. 
\textbf{(b)} Methodology overview, outlining the semi-structured interview protocol with guiding questions and the primary themes of analysis derived from the responses. 
\textbf{(c)} Sector-specific analysis showing the percentage of interviewees within each sector who affirmed four key propositions: the urgency of AI misuse in biology, the inadequacy of current screening protocols, the need for governance, and support for functional screening. 
\textbf{(d)} Co-occurrence matrix of key themes across all 130 interviewees. The values indicate the number of individuals who hold both intersecting views.}
\label{fig:interview_summary}
\end{figure}
To understand whether this gap between AI's capabilities and existing safeguards is a theoretical concern or a present danger, we went directly to the experts. We conducted in-depth interviews with 130 stakeholders across industry, government, academia, and policy to systematically assess their views on emerging threats and the sufficiency of current defenses. The interviewees were distributed across four key sectors (Figure \ref{fig:interview_summary}a): Industry (n=43), including representatives from nucleic acid synthesis companies, screening technology providers, and AI companies; Academia (n=41), with researchers from institutions like MIT, Stanford, Harvard, and Lawrence Livermore National Laboratory; Government (n=29), with personnel from the Department of Homeland Security, National Institutes of Health, and the Federal Bureau of Investigation; and Policy (n=17), including leaders from organizations like the American Association for the Advancement of Science and the Engineering Biology Research Consortium. The semi-structured interviews, lasting 30-60 minutes each, used guiding questions focused on emerging risks, the efficacy of current safeguards, and gaps in existing governance frameworks (Figure \ref{fig:interview_summary}b).
Across these interviews, four key themes consistently emerged, as shown in Figure \ref{fig:interview_summary}c:
\begin{itemize}
    \item \textbf{AI Misuse is an Urgent Concern in Bio:} 76\% of participants highlighted the urgency of AI misuse in biological domains. Concern was particularly pronounced among policy experts (94\%) and academics (83\%).
    \item \textbf{Calls for Governance:} 74\% of participants advocated for clearer governance standards to address emerging threats. This call was nearly universal among policy stakeholders (94\%) and strongly supported by academia (85\%).
    \item \textbf{Current Screening Tools are inadequate:} 47\% of interviewees expressed skepticism about existing sequence-based screening systems. This sentiment was strongest in the policy (82\%) and academic (59\%).
    \item \textbf{Supports Functional Screening:} A majority of experts supported the development of functional screening methods as a necessary supplement to sequence-based approaches, with especially strong endorsement from government and industry representatives.
\end{itemize}

Interviewees frequently linked these concerns, revealing a broader pattern of interconnected priorities (Figure \ref{fig:interview_summary}d). Among interviewees who emphasized the urgency of AI misuse, 91.8\% also highlighted inadequacies in current screening methods, 90.8\% advocated for functional screening approaches, and 80.2\% supported stronger governance measures. These patterns suggest that concern over AI-driven biosafety risks is not isolated but rather reflects a more comprehensive recognition that current biosecurity infrastructure is inadequate to meet emerging threats. 83.6\% of interviewees viewed current screening tools as inadequate, while also supporting stronger governance standards and endorsing functional screening approaches. This covariance analysis reveals strong positive correlations among these three attitudes, offering quantitative evidence of systematic alignment across diverse stakeholders in support of comprehensive policy and technological reforms.

Beyond broad thematic alignment, interviews identified distinct operational pressures and governance challenges unique to each domain. Academic researchers emphasized the need for low-friction safeguards that integrate seamlessly into scientific workflows and avoid flagging clearly legitimate requests (e.g., an authorized Ebola researcher ordering Ebola strains). Regulatory gray zones, such as gain-of-function work in BSL-1/2 labs, complicate institutional accountability and raise questions about the consistency of oversight. Research agencies like the NIH face the delicate task of distinguishing legitimate science from dual-use risk without stifling innovation. Investigative bodies such as the FBI, meanwhile, must navigate a rapidly evolving threat landscape, and called for tools with greater interpretability and clearer confidence metrics to support real-time decision-making.

On the industry side, nucleic acid synthesis providers operate under uneven incentives and technical constraints. Some actively screen for hazardous sequences, motivated by safety culture, liability exposure, and reputational risk. Those who opt out cite high costs, technical limitations, and operational complexity. Screening providers are themselves divided: some acknowledge the emerging risk of AI-edited pathogens, yet remain skeptical that truly \textit{de novo} threats are imminent, citing current technological limits. Cloud labs add another layer of complexity: while enabling rapid, high-throughput experimentation, they often fall outside the International Gene Synthesis Consortium (IGSC) frameworks, introducing oversight blind spots as synthesis and experimentation become increasingly abstracted from traditional governance structures.

Our survey revealed a unifying message: biosecurity professionals across sectors are calling for next-generation safeguards capable of keeping pace with the accelerating capabilities of AI-enabled biology.
\begin{Takeaway}{The Message is Clear}{}
\textbf{Across academia, government, industry, and policy, one message resounded:} Today’s screening tools and governance structures are no longer sufficient. Without functional screening, secure-by-design GenAI architectures, and harmonized global standards, the pace of AI innovation risks exceed our biosafety infrastructure.
\end{Takeaway}

\section{A Roadmap for Safe and Secure GenAI in Biosciences}

\noindent
Securing GenAI in the biosciences is not about finding a single solution. Instead, it requires building a fortress with multiple layers of defense, each designed to anticipate, withstand, and adapt to threats. This roadmap outlines interventions across three critical stages of a GenAI model’s lifecycle: from the data that builds it (pre-training), to the rules that shape it (post-training), to the guards that protect it in action (inference), as shown in Figure. \ref{fig:biosafe_framework}. While many of these techniques originate in the AI safety literature \cite{liu2023trustworthy, chua2024ai}, this section translates them into accessible principles for biosafety practitioners, policymakers, and medical stakeholders. By building these layers of defense, we can create a resilient infrastructure that fosters innovation while protecting against catastrophic risk.

\paragraph{Laying the Foundation: Securing AI’s base (Pre-training Stage):} Every strong fortress begins with carefully chosen materials. For AI, this means selecting, filtering, and securing the data it learns from. If unsafe biological knowledge is built into the foundation, future safeguards may never catch it. This stage is about keeping dangerous knowledge out of the system from the start.


\begin{itemize}[leftmargin=1.5em]
    
    \item \textbf{Blueprint Control: Access Restrictions.} Preventing the misuse of GenAI in biosciences begins with stringent access control over sensitive biological data. Genomic sequences of high-risk pathogens, toxin-coding regions, and gene drive constructs represent a category of “blueprint data” that, if incorporated into model training without proper oversight, could significantly elevate dual-use risks. To mitigate such threats, access to these datasets should be governed by role-based permissions, comprehensive usage logging, and rigorous biosafety risk classifications. However, safety considerations must extend beyond data access and encompass downstream model availability as well. Models trained on sensitive datasets should be subject to restricted access protocols, wherein distribution of model weights, inference APIs, or fine-tuned checkpoints is contingent on identity verification, institutional affiliation, or formal approval by governing entities. This layered approach, applying control both upstream at the data level and downstream at the deployment interface, ensures that biological foundation models remain aligned with safety and security mandates throughout their lifecycle.

    \item \textbf{Toxic Material Checks: Dataset Filtering.} A cornerstone of AI biosafety is rigorous data-centric curation, aimed at minimizing the model’s exposure to harmful or dual-use biological knowledge during both training and fine-tuning. This begins with a process of \textit{dataset risk stratification}, in which all candidate training sequences are systematically screened using tools such as \textit{BLAST} against established databases of known pathogens, toxins, or sequences of concern. Sequences flagged as high-risk are then either excluded entirely or selectively obfuscated to prevent the model from memorizing and reproducing hazardous content. This principle extends to \textit{synthetic data augmentation}, where any artificially generated sequences, often used to increase dataset diversity, must undergo the same level of scrutiny. Computational safety filters can be employed to evaluate properties such as toxicity or similarity to restricted sequences, ensuring that only benign and policy-compliant data are incorporated during retraining or model updates. By embedding these safeguards at both ends of the data pipeline, natural and synthetic, dataset filtering forms a foundational defense layer that reduces the likelihood of a model inadvertently acquiring the ability to generate dangerous biological constructs.

        \item \textbf{Manufacturer’s Stamp: Watermarking for Traceability. } One promising approach to ensuring accountability in AI-enabled biosciences is the use of \textit{watermarking}, embedding subtle, traceable signatures into model-generated biological outputs. In the context of bio-foundation models, such watermarks could be applied to synthetic DNA sequences, protein structures, or molecular designs to indicate their origin. These signatures, whether cryptographic or statistical in nature, offer a mechanism to trace a given output back to its model version, training dataset, or point of generation. This is especially important in dual-use contexts, where it may be critical to identify whether a harmful agent was designed using open-source AI tools or proprietary systems.

The concept draws inspiration from watermarking research in the large language model (LLM) community \cite{kirchenbauer2024watermarklargelanguagemodels, wmark1, wmark2, wmark3, wmark4, wmark5}. Recent methods partition a model’s vocabulary into controlled subsets—biasing generation toward statistically identifiable token patterns that remain invisible to the human reader. In the field of biology, FoldMark \cite{foldmark} makes a pioneering attempt to embed traceable watermarks into protein structures. According to the paradigm of FoldMark, it proposed embedding watermarks into protein structures using a two-stage process: first, training a dedicated encoder to insert a robust signature into protein geometry, and second, fine-tuning a generative model to preserve this signature during molecule creation. These approaches aim to create “manufacturer’s stamps” within the fabric of biological outputs, akin to serial numbers on lab equipment, enabling downstream detection and forensic analysis.

\begin{figure*}[h!]
	\centering
  \includegraphics[width=0.9\linewidth]{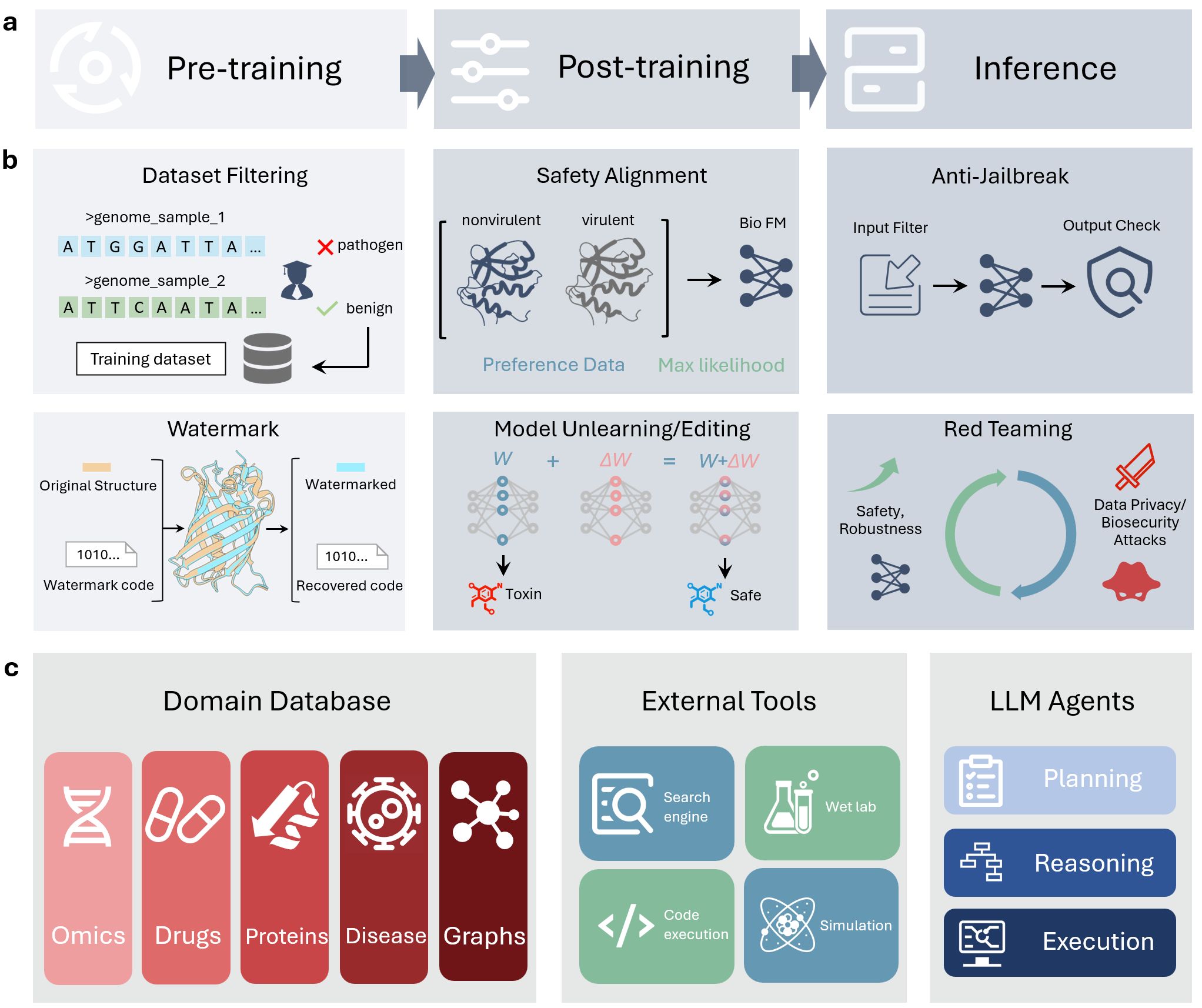}
    \caption{\textbf{Towards Safe and Secure GenAI in Biosciences}
    \textbf{(a)} The framework operates across the three stages of a model's lifecycle: pretraining, finetuning, and inference.
    \textbf{(b)} Internal safety strategies deployed by BioSafe, including dataset filtering, safety alignment, anti-jailbreak checks, watermarking, model unlearning, and continuous red teaming.
    \textbf{(c)} The agent's capabilities are powered by access to domain databases, external tools like search and simulation, and its core LLM architecture for planning, reasoning, and execution.
    }
    \label{fig:biosafe_framework}
	\label{figure4}
\end{figure*}

However, watermarking is not a silver bullet. In the text domain, adversaries can already strip or obfuscate watermarks through paraphrasing, compression, or adversarial re-generation. Similar evasion techniques may emerge in bio domains, e.g., altering protein folding or substituting synonymous codons to bypass sequence-based detectors. This highlights a broader challenge: watermarking can offer a layer of accountability, but it must be complemented by additional safeguards such as access controls, metadata provenance, and usage monitoring. Further interdisciplinary research is urgently needed to adapt and harden watermarking systems for the unique challenges of biosecurity.

    \end{itemize}

\begin{tcolorbox}[colback=gray!5!white,colframe=gray!80!black,title=Key Takeaway]
Even the most advanced AI models are only as safe as the data they learn from. Pre-training safety is about keeping hazardous knowledge out before it ever enters the system.
\end{tcolorbox} 

\paragraph{Training the Guardian: Teaching AI What Not to Do (Post-training Stage): } Once built, the model must be shaped into a responsible agent. This stage is about instilling guardrails through feedback, stress testing, and targeted forgetting. Like training a powerful soldier, the goal is to teach the GenAI what to do and what never to do.

\begin{itemize}[leftmargin=1.5em]
    \item \textbf{Moral Training: Safety Alignment via RLHF.}
To ensure the safe deployment of generative models in biosciences, alignment techniques must embed ethical, regulatory, and biosecurity considerations directly into model behavior. Reinforcement Learning with Human Feedback (RLHF) and supervised fine-tuning have emerged as leading strategies to train models toward reliable and responsible outputs~\cite{dai2023safe, rlhf1, rlhf2, rlhf3, rlhf4, rlhf5}. These techniques enable models to internalize safety-relevant preferences by incorporating feedback from domain experts and human evaluators, guiding the generation process in accordance with public health and biosecurity standards. 
In practice, the RLHF process unfolds in three key stages. First, supervised fine-tuning (SFT) adapts a pre-trained model using curated examples of safe and desirable outputs, forming an initial alignment with normative standards. Second, a reward model (RM) is trained, typically using pairwise comparisons of outputs, to quantify how well each response adheres to intended goals. Finally, reinforcement learning is used to optimize the generative model so as to maximize the expected reward while constraining deviation from the original model. This iterative refinement has proven effective in large-scale systems such as ChatGPT, Gemini, and Claude, where RLHF serves as a cornerstone of safety and value alignment.

However, applying RLHF to GenAI presents novel challenges. Unlike natural language, biological sequences, such as proteins, RNAs, or regulatory DNA regions, lack intuitive semantics and are often unreadable to non-experts. Their meaning is deeply embedded in structure-function relationships that are difficult to evaluate without computational tools or wet-lab validation. As a result, designing reward functions for bioscience applications requires careful integration of safety constraints with biological validity and functional correctness.
To address these domain-specific hurdles, alignment frameworks can be enhanced by incorporating traditional bioinformatics methods such as sequence alignment, structural similarity scoring, or known detection into the reward modeling process. These tools offer interpretable signals for biological plausibility and can guide models away from generating potentially harmful or non-functional sequences. Additionally, recent work on aligning biological languages with natural language representations~\cite{de2025multimodal} suggests a promising transfer pathway: applying the RLHF paradigm developed for large language models to restrict biologically dangerous outputs by grounding them in interpretable natural language cues—particularly useful for therapeutic or disease-associated applications.

\medskip

\noindent
Overall, adapting RLHF for bio-generative models offers a compelling path forward, but one that demands domain-specific innovation, interpretability tools, and deep collaboration between biologists, AI researchers, and safety experts.
During finetuning, models can be aligned with explicit safety, ethical, and regulatory frameworks. Techniques based on Supervised fine-tuning and RLHF \cite{dai2023safe, rlhf1, rlhf2, rlhf3, rlhf4, rlhf5} have emerged as methods to iteratively guide model outputs toward safe, reliable behavior by training the model parameters. With feedback from domain experts and human evaluators, the model can learn to prioritize outputs aligned with public health and biosecurity standards.

Adversarial training has proven effective in both natural language and computer vision domains~\cite{bai2021recent, adv1, adv2, adv3, adv4}, significantly improving model robustness against real-world attacks. Translating this paradigm to biological foundation models presents a unique opportunity: by incorporating expert-curated unsafe prompt corpora and generating diverse misuse scenarios specific to the life sciences, developers can train models that recognize and resist biological threat vectors. This proactive approach strengthens the model's ability to generalize beyond known threats and provides a critical layer of resilience as generative capabilities continue to expand in the biosciences.

    \item \textbf{Memory Surgery: Model Unlearning.}
Model unlearning refers to the process of selectively removing sensitive, harmful, or outdated knowledge from a trained model, without retraining it entirely from scratch. Originally developed in the context of large language and vision-language models, unlearning has emerged as a key technique for mitigating the retention of toxic, biased, or dual-use information embedded during training. This is typically accomplished through methods such as gradient ascent, knowledge editing, or negative preference optimization, which increase the model’s uncertainty around specific undesirable outputs while preserving general capabilities across safe examples.
Conceptually, given an input prompt $x$ that leads a model $f(x)$ to produce a harmful output $y_{\text{danger}}$, the goal of unlearning is to disrupt this association, suppressing $y_{\text{danger}}$ without impairing the model’s ability to generate appropriate outputs for benign inputs. In the biological domain, this technique is especially relevant for foundation models that may inadvertently memorize or reproduce hazardous sequences, such as genes encoding lethal toxins or virulence factors. As dual-use risks often emerge only after deployment, the ability to surgically erase dangerous content becomes essential for maintaining long-term safety.

To operationalize unlearning in bioscience models, researchers have proposed techniques that fine-tune models against “negative examples” or apply targeted gradient updates to reduce recall of specific biological sequences. For instance, increasing the loss, or perplexity, on known sensitive patterns causes the model to effectively “forget” this knowledge, rendering it less accessible at inference time. Recent studies in large language models~\cite{Blanco_Justicia_2025} have shown that such interventions can be both efficient and precise. Key insights include the value of gradual forgetting (to avoid catastrophic interference), targeting short subsequences rather than broad content spans, and maintaining output consistency on safe prompts to ensure model utility is preserved. These findings offer a promising foundation for adapting unlearning to biological contexts, where statistical and structural parallels with natural language can be leveraged to guide implementation. 

\end{itemize}
\begin{tcolorbox}[colback=gray!5!white,colframe=gray!80!black,title=Policy Insight]
Techniques like RLHF and adversarial training teach models to recognize and avoid misuse scenarios, but safety is never guaranteed. Clear standards for when and how to apply them are needed.
\end{tcolorbox}

\paragraph{Guarding the Gates: Real-Time Defense During Use (Inference Stage): }
Even with careful training, no model is perfect. That’s why active defenses are needed during deployment. This layer monitors user interactions and generates content in real-time, like guards stationed at the gates of the fortress.

\begin{itemize}[leftmargin=1.5em]

\item \textbf{The Front Checkpoint: Anti-Jailbreak Screening.} A comprehensive defense against model misuse requires safeguards at both the input and output levels. At the input level, the primary goal is to manage user prompts to prevent malicious queries from triggering the generation of harmful content. This is achieved through a pre-screening module that combines \textbf{prompt classification} to block dangerous requests and \textbf{prompt optimization} to rewrite potentially risky queries into safer variants. This defensive layer uses a mix of traditional bioinformatics tools (e.g., \textit{BLAST}) for known threats and deep learning classifiers for novel or obfuscated prompts, securing the model at the crucial user-input interface.

\par At the output level, safety is reinforced by a critical screening and filtering layer designed to detect and block any hazardous biological sequences that may still be generated. This involves a two-pronged approach. First, traditional \textbf{rule-based filtering} uses homology-based tools like \textit{BLAST} to flag outputs that show high sequence similarity to known pathogens or toxins. To address the limitations of this method against novel threats, this is augmented with advanced \textbf{function-based screening}. Tools like Omnyra \cite{H4D_Omnyra} leverage protein language models to assess a sequence's potential functional risk, providing a more future-proof defense by focusing on what a sequence might do rather than what it looks like.

    \item \textbf{Live Fire Drills: Red-Teaming for Vulnerability Discovery.} Red-Teaming employs specific strategies to design diverse prompt inputs, enabling the model to generate potentially harmful content under controlled conditions. This approach aims to uncover vulnerabilities in the model that could lead to undesirable behavior. In Red-Teaming tests for large language models \cite{perez2022red}, common strategies include the use of technical slang, reframing prompts, authority manipulation, and even the inclusion of garbled prefixes may work. The outcomes of these tests provide valuable insights to developers, assisting them in considering and implementing security enhancements for the model. Furthermore, this methodology is also worth to be applied to the security evaluation of bio foundation models. Continuous stress testing by interdisciplinary red-teaming efforts—bringing together experts in biology, machine learning, cybersecurity, and ethics—helps uncover novel vulnerabilities and emergent misuse pathways. These teams simulate attack scenarios, test model defenses, and identify gaps that may otherwise go unnoticed. The insights gained support the deployment of adaptive countermeasures, ensuring the model remains secure and robust as threat landscapes evolve.

    \item \textbf{Autopilot Override: Inference-Time Alignment.} Inference-time alignment offers a flexible approach to steer bio-foundation model outputs toward safety without retraining. In white-box settings where model logits are accessible, \textbf{controlled decoding} \cite{cd, transferq, args, immune} modifies token-level probabilities during generation. By integrating domain-specific evaluators—such as toxicity predictors or pathogen classifiers \cite{pathollm}—this method can down-weight unsafe tokens in real-time to guide generation toward biologically safe and compliant sequences. When direct access to logits is unavailable (i.e., in \textbf{black-box} scenarios), alignment is instead achieved by generating and evaluating full candidate sequences. Common strategies include \textbf{parallel sampling} (e.g., Best-of-N) \cite{bon1, bon2, bon3}, which generates multiple outputs and selects the one that scores highest on a safety evaluator, and \textbf{sequential refinement} \cite{refine1, pair, flirt, feedback}, which iteratively improves a response using evaluator feedback. Together, these techniques provide a crucial safety layer adaptable to different model access levels, enabling the responsible deployment of powerful bio-foundation models.   

\end{itemize}

\begin{tcolorbox}[colback=gray!5!white,colframe=gray!80!black,title=Real-World Readiness]
Inference-stage tools are the last line of defense. They must be fast, adaptable, and deeply integrated with biological risk models, not just keyword filters.
\end{tcolorbox}

\paragraph{A Living Fortress: Toward Adaptive Biosafety and Biosecurity Systems.} Securing the future of GenAI in the biosciences will require more than static safeguards, it demands infrastructure that is intelligent, integrated, and adaptive. As GenAI models become more capable of planning, simulating, and generating biological designs, our oversight systems must evolve in lockstep, offering not just filters but informed, real-time risk assessments and dynamic intervention strategies.

\smallskip

\noindent
\textbf{Integration with domain knowledge.} Next-generation biosafety systems will need to interface deeply with biological knowledge. This includes connecting to curated databases of genomic sequences, protein functions, pathogenic mechanisms, and therapeutic compounds. Such integration enables contextual analysis—allowing safety layers to simulate downstream effects, verify biological plausibility, and anticipate potential harms with domain-specific grounding. These tools must be paired with external capabilities such as molecular simulation engines, structure predictors, and API-accessible regulatory frameworks to validate emerging risks at inference time.

\smallskip

\noindent
\textbf{Towards agentic capabilities.} Beyond knowledge integration, emerging safeguards must also demonstrate higher-order reasoning: the ability to plan multi-step responses, evaluate scientific and policy constraints, and red-team outputs against evolving threats. Drawing inspiration from recent advances in agentic AI, these systems could modularly score dual-use potential, assess confidence in high-stakes generations, and trace prompt-output pathways for auditability. To remain trustworthy, their actions should be explainable, logged, and subject to human oversight.

\smallskip

\noindent
\textbf{Self-Evolving Capabilities.} Finally, such oversight mechanisms must themselves be capable of evolution. The biosafety tools of the near future must learn from new threat patterns, incorporate feedback from red-team exercises, and stay synchronized with technological progress—whether that’s the release of more powerful models or the emergence of novel biological capabilities like virtual cell simulation. An effective biosafety system is not a firewall, it is a living guardian.

\begin{tcolorbox}[colback=blue!5!white,colframe=blue!75!black,title=Final Message]
No single measure can fully secure GenAI in the biosciences. But layered interventions spanning training, deployment, access, and monitoring can collectively build a resilient infrastructure to safeguard GenAI. Security in this domain is not static policy, but dynamic practice. And staying ahead requires oversight mechanisms and technical capabilities and evaluation tools that evolve as fast as the GenAI models they guard.
\end{tcolorbox}




\section{Conclusion: From Awareness to Action}
This Perspective has surfaced a growing expert consensus: GenAI models operating in the biosciences present unprecedented biosecurity risks that outpace the assumptions of existing safeguards. Through over 130 interviews with global leaders in synthetic biology, cybersecurity, governance, and AI development, we have highlighted the dissonance between the old playbook designed for physical gene synthesis and the new frontier, where models can design dangerous biological sequences without ever ordering DNA. The risks are no longer theoretical. The tools are widely accessible. And the gaps in oversight are widening.

Addressing this rapidly shifting threat landscape requires a fundamental rethinking of how we govern dual-use technologies. First, technical safeguards must be embedded across the AI lifecycle from training-time dataset curation to inference time filtering, access controls, output watermarking, and red-teaming protocols. These tools must be explainable, auditable, and grounded in biological domain knowledge. Second, interdisciplinary collaboration is critical: no single stakeholder group can anticipate or manage these risks alone. Biologists, AI researchers, ethicists, national security experts, and policymakers must work together to build a resilient safety architecture. Third, we need adaptive and strong governance and policies that may include tiered access, model registration, risk-based release standards, and international coordination.

Ultimately, securing the bio-AI frontier is not a matter of technical patchwork or post-hoc policy. It requires a layered, dynamic, and forward-looking strategy, one that evolves as fast as the models it seeks to protect. Only through anticipatory, cross-sectoral, and globally coordinated action can we harness the transformative potential of GenAI in the life sciences while minimizing the profound risks it brings.

\bibliography{main}

\clearpage

\section*{Acknowledgements}

\section*{Author contributions statement}

\section*{Competing interests}
The authors declare no competing interests.
Disclaimer: Certain tools and software are identified in this paper to foster understanding. Such identification does not imply recommendation or endorsement by the National Institute of Standards and Technology, nor does it imply that the tools and software identified are necessarily the best available for the purpose.
\section*{Additional information}

{\bf Correspondence and requests for materials} should be addressed to Mengdi Wang.

\clearpage
\setcounter{figure}{0}

\clearpage
\setcounter{figure}{0}
\setcounter{table}{0}
\makeatletter 
\renewcommand{\thefigure}{S\@arabic\c@figure}
\renewcommand{\thetable}{S\@arabic\c@table}
\makeatother
\appendix

\section{GenAI in Biosciences}

\begin{table*}[h]
\centering
\label{tab:frontier_models}
\begin{tabular}{|p{3.5cm}|p{6.5cm}|p{6.5cm}|}
\hline
\textbf{Biological Target} & \textbf{Frontier AI Models} & \textbf{Purpose / Capability} \\
\hline
Proteins & ESM, ESM-2, ESM-3, Progen, Progen2, ProteinMPNN, RFdiffusion, RFdiffusionAA, PocketGen, BindCraft & Learn protein sequence–structure–function relationships; predict structures; generate novel proteins; design functional protein backbones and ligand-binding sites. \\
\hline
RNA & RNAGenesis, RNA-FM & Model RNA sequence–structure relationships; predict secondary structures; annotate functions; enable de novo RNA design. \\
\hline
DNA & Nucleotide Transformer, Evo, Evo2 & Model genomic sequences at varying scales; detect functional elements; enable long-context genome modeling and design. \\
\hline
Small Molecules & JT-VAE, DeLinker, SyntheMol, MolDiff, Uni-Mol & Explore large chemical space; design bioactive, synthesizable molecules; model 2D/3D molecular structures and binding interactions. \\
\hline
Single Cells & scBERT, scFoundation, Geneformer, scDiffusion & Learn cell-type representations; correct batch effects; predict perturbation responses; simulate transcriptomic profiles and developmental trajectories. \\
\hline
Clinical / Medical Data & CHIEF, CONCH, MINIM & Analyze histopathology images; predict cancer origin and prognosis; integrate image–text understanding for clinical interpretation. \\
\hline
\end{tabular}
\caption{Generative AI models across different biological scales and their primary capabilities.}
\end{table*}

\newpage

\section{Explanation of Terms}

\begin{table}[h]
\centering
\begin{tabular}{|p{4cm}|p{10cm}|}
\hline
\textbf{Term} & \textbf{Definition} \\ 
\hline
Pathogen & A biological agent that can cause disease in its host, such as bacteria, viruses, fungi, or parasites. \\ \hline
Toxin & A poisonous substance produced by living organisms that can cause disease or death when introduced into the body. \\ \hline
Omics & A branch of biology that studies the complete set of a particular type of biomolecule or molecular process in an organism, such as genomics (genome), proteomics (proteome), and metabolomics (metabolome). \\ \hline
Genome & The complete set of genetic material in an organism, typically referring to nuclear DNA in eukaryotes, but can include other DNA such as mitochondrial or chloroplast DNA. \\ \hline
Nucleic acids & Macromolecules that store and transmit genetic information in living organisms, composed of nucleotides, which consist of a sugar, a phosphate group, and a nitrogenous base. The two main types are DNA and RNA. \\ \hline
CRISPR & A gene-editing technology that uses the CRISPR-Cas9 system to make precise changes to the DNA of living organisms, based on a natural defense mechanism in bacteria. \\ \hline
Embryo & The early stage of development of a multicellular organism, typically from fertilization until the end of the eighth week in humans, after which it is called a fetus. \\ \hline
Bioweapon & A type of weapon that uses biological agents, such as bacteria, viruses, or toxins, to cause disease or death in humans, animals, or plants. \\
\hline
\end{tabular}
\caption{Glossary of key biological terms}
\end{table}

\begin{longtable}{|p{5cm}|p{10cm}|}

\hline
\textbf{Term} & \textbf{Definition} \\ \hline
\endfirsthead

\hline
\textbf{Term} & \textbf{Definition} \\ \hline
\endhead

\hline
\multicolumn{2}{|r|}{{Continued on next page}} \\ \hline
\endfoot

\hline

\caption{Glossary of Key Machine Learning Terms} \label{tab:glossary} \\
\endlastfoot

Foundation Models & Large deep learning models pre-trained on vast, general-purpose datasets, designed to be versatile and fine-tuned for specific applications. \\ \hline
Deep Generative Models & Machine learning models that use deep neural networks to generate new data similar to their training data. \\ \hline
Fine-Tuning & The process of adapting a pre-trained model to a new task by further training it on a new dataset, especially useful when the new dataset is small. \\ \hline
Reinforcement Learning & A machine learning paradigm where an agent learns to maximize a reward signal by interacting with an environment. \\ \hline
Retrieval-Augmented Generation & A technique that enhances large language models by providing them with external information to improve the accuracy and relevance of their generated text. \\ \hline
Self-supervised Learning & A machine learning method where a model learns from unlabeled data by generating its own labels or supervisory signals. \\ \hline
AI Agent & A software system that uses artificial intelligence to perform tasks or make decisions based on input prompts. \\ \hline
Multi-Agent System & A system consisting of multiple AI agents that interact with each other and their environment to achieve individual or collective objectives. \\ \hline
Jailbreak Attacks & Techniques used to manipulate AI models, particularly large language models, into producing incorrect or fabricated information that circumvents their built-in safety mechanisms. \\ \hline
Membership Inference Attacks & Attacks where an adversary tries to determine whether a specific data record was used to train a machine learning model. \\ \hline
Watermark & In AI, a technique to mark or identify content generated by an AI model, often used to verify the origin or authenticity of the content. \\ \hline
Surrogate Model & A simpler model used to approximate the behavior of a more complex or computationally intensive model, often for interpretability or efficiency reasons. \\ \hline
Multi-Modality Model & A machine learning model that can process and integrate information from multiple different types of data, such as text, images, and audio. \\ \hline
Parameter Efficient Finetuning & Techniques for fine-tuning large pre-trained models by updating only a small subset of their parameters, to make the process more computationally efficient. \\ \hline
Prompt & The input text or instruction provided to an AI model, particularly large language models, to guide its output. \\ \hline
Hallucination & When an AI model, especially a large language model, generates incorrect or fabricated information that is presented as if it were true. \\ \hline
Transformers & A type of neural network architecture that uses self-attention mechanisms to process sequences, such as text, in parallel and understand the relationships between different parts of the sequence. \\ \hline
Embodied AI & Artificial intelligence systems that are integrated into physical entities, like robots, to interact with and learn from their environment through sensors and actuators. \\ \hline
Backdoor Attacks & Attacks where an adversary modifies a machine learning model during its training phase to behave in a specific, malicious way when presented with certain inputs or triggers. \\ \hline
Property Inference Attacks & Attacks where an adversary tries to infer specific properties or characteristics of the training data used to train a machine learning model, by analyzing the model's behavior or outputs. \\ \hline

\end{longtable}
\clearpage

\section{Survey Methodology}

We led an in-depth qualitative survey of stakeholders across industry, government, academia, and policy to highlight emerging biosecurity concerns at the intersection of artificial intelligence, synthetic biology, and governance. Over a ten-week period, we conducted 130 semi-structured discovery interviews with individuals spanning four key sectors:
\begin{itemize}
  \item 43 professionals from private sector companies including nucleic acid synthesis companies, screening technology providers, and technology and AI companies. 
  \item 41 academics and researchers spanning disciplines such as microbiology, virology, machine learning, and synthetic genomics. Institutions represented included the Lawrence Berkeley National Laboratory, Lawrence Livermore National Laboratory, Stanford Biosecurity, MIT Media Lab, Wyss Institute at Harvard University, and the Gladstone Institutes.
  \item 29 representatives from government agencies, including public health authorities, national security offices, and regulatory bodies. These organizations included the Department of Homeland Security, National Institutes of Health, Federal Bureau of Investigation, U.S. Navy, National Security Commission on Emerging Biotechnology, National Institute of Standards and Technology, and the Defense Innovation Unit.
  \item 17 non-governmental policy experts and think tank leaders specializing in biosecurity advocacy and AI governance. Organizations represented included the Federation of American Scientists, American Association for the Advancement of Science,  Engineering Biology Research Consortium, and Centre for Long-Term Resilience.

\end{itemize}
Interviewees were selected to reflect a diversity of institutional roles, disciplinary backgrounds, and viewpoints, with many recognized as experts in their respective fields. Each interview lasted between 30 to 60 minutes and was conducted using video conference. Interviews followed a semi-structured protocol, beginning with open-ended prompts about current and future biosecurity risks and then progressing organically based on the interviewee’s expertise and perspectives. 
To ensure consistency across interviews while allowing flexibility, we used a core set of guiding questions as listed below: 
\begin{itemize}
    \item What emerging risks or concerns do you anticipate at the intersection of artificial intelligence and synthetic biology?
    \item How effective do you think current biosecurity practices are in addressing AI-enabled threats?
    \item What gaps, if any, do you see in the existing tools, frameworks, or governance systems for detecting and preventing misuse in the synthetic biology context?
    \item What types of interventions do you think are most needed to strengthen safeguards in biosecurity?
    \item What role should your sector play in shaping the future of biosecurity as capabilities evolve?
\end{itemize}
To analyze and quantify interview responses, we developed a framework around four key themes: (1) AI misuse in biological contexts, (2) perceptions of current sequence screening tools, (3) attitudes toward functional or AI-native screening approaches, and (4) governance gaps and regulatory needs. Each interview was reviewed and assigned a binary relevance score (1 = substantial mention, 0 = no substantial mention) for each theme. A theme was coded as “1” only if the interviewee offered a substantive comment that demonstrated support, engagement, or direct relevance to the topic such as describing concrete practices, expressing informed perspectives, or endorsing the importance of the issue. Passing mentions or unrelated concerns were not counted.

This framework allowed us to quantify the prevalence of each theme across all 130 interviews, compare thematic engagement across stakeholder groups, and identify patterns in concern and/or alignment across sectors.

\section{Interview Results}

\begin{table}[H]
\centering
\caption{Summary of Interview Themes by Sector}
\begin{tabular}{lcccc}
\hline
\textbf{Sector} & 
\makecell[c]{\textbf{AI Misuse}\\\textbf{is Urgent}} & 
\makecell[c]{\textbf{Screening Tools}\\\textbf{are Inadequate}} & 
\makecell[c]{\textbf{Supports Functional}\\\textbf{Screening}} & 
\makecell[c]{\textbf{Calls for Governance}\\\textbf{Standards}} \\
\hline
Industry & 30 & 10 & 25 & 24 \\
Government & 19 & 13 & 15 & 21 \\
Academia & 34 & 24 & 24 & 35 \\
Policy & 16 & 14 & 12 & 16 \\
\hline
\textbf{Total} & \textbf{99} & \textbf{61} & \textbf{76} & \textbf{96} \\
\hline
\end{tabular}
\label{tab:theme_summary}
\end{table}

\end{document}